# Designing Corrosion-Resistant CoCrNi Medium Entropy Alloys via Short-Range Order Modification

Elaf A. Anber[1#], Debashish Sur[2,3#], Annie K. Barnett[1], Daniel L. Foley[1], Andrew M. Minor[4], Brian DeCost[5], Howie Joress[5], Anatoly I. Frenkel[6,7], Michael L. Falk[1,8,9,10], John R. Scully[2,3], Mitra L. Taheri[1*]

[1]*Department of Materials Science & Engineering, Johns Hopkins University, Baltimore, MD 21218 USA*
[2] *Center for Electrochemical Science and Engineering, University of Virginia Charlottesville VA 22904 USA*
[3]*Department of Materials Science and Engineering, University of Virginia Charlottesville VA 22904 USA*
[4]*Department of Materials Science and Engineering, University of California-Berkeley and National Center for Electron Microscopy, Molecular Foundry, Lawrence Berkeley National Laboratory, CA 94720 USA*
[5]*Materials Measurement Science Division, National Institute of Standards and Technology, Gaithersburg, MD USA*
[6] *Department of Materials Science and Chemical Engineering, Stony Brook University, Stony Brook, NY 11794, USA.*
[7] *Chemistry Division, Brookhaven National Laboratory, Upton, NY 11973, USA.*
[8]*Department of Mechanical Engineering, Johns Hopkins University, Baltimore, MD 21218 USA*
[9]*Department of Physics and Astronomy, Johns Hopkins University, Baltimore, MD 21218 USA*
[10]*Hopkins Extreme Materials Institute, Johns Hopkins University, Baltimore, MD 21218 USA*

# *Authors contributed equally*
*Corresponding Author: Mitra Taheri* mtaheri4@jhu.edu.

## Abstract

Equiatomic CoCrNi medium entropy alloys are known for their unique properties linked to chemical short-range order (CSRO), crucial in both percolation processes and/or nucleation and growth processes influencing alloy passivation in aqueous environments. This study combines extended x-ray absorption fine structure, atomistic simulations, electrochemical methods, x-ray photoelectron spectroscopy, and transmission electron microscopy to explore CSRO evolution, passive film formation, as well as its characteristics in the as-homogenized CoCrNi condition, both before and after aging treatment. Results reveal a shift in local alloying element bonding environments post-aging, with simulations indicating increased Cr-Cr ordering in 2$^{nd}$ nearest neighbor shells. Enhanced passive film formation kinetics and superior protection of the aged alloy in harsh acidified 3 mol/L NaCl solution indicate improved aqueous passivation correlated with Cr-Cr CSRO. This work establishes a direct connection between alloy CSRO and aqueous passivation in CoCrNi, highlighting the alloy's potential for tailored corrosion-resistant applications.

**Graphical Abstract**

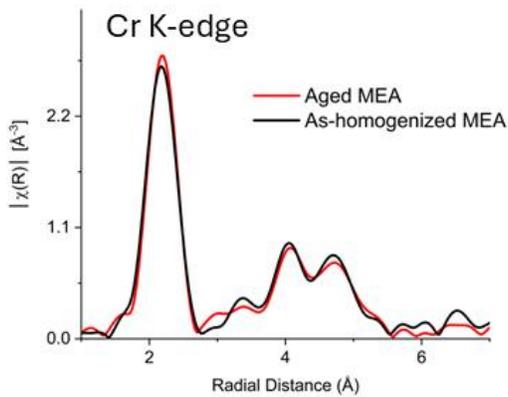
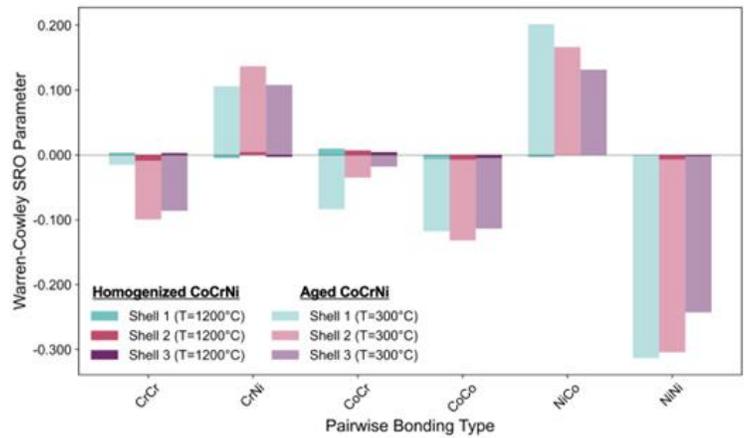
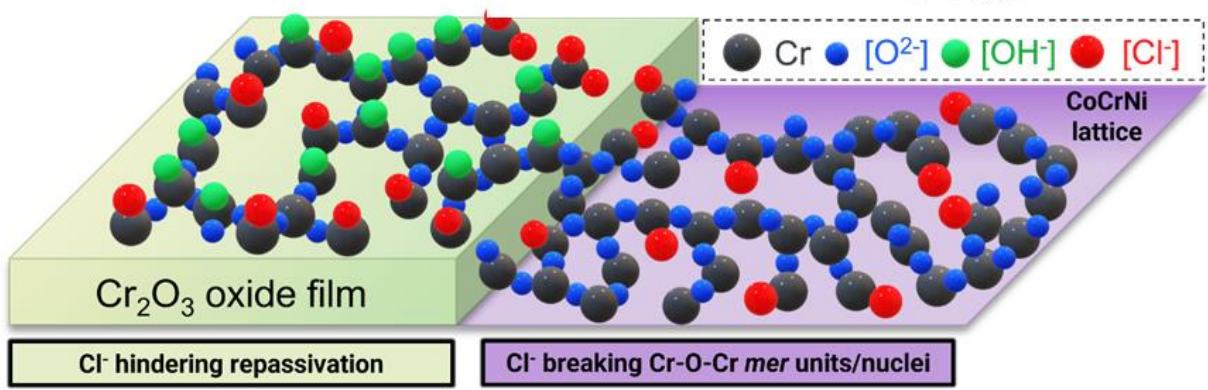

**Introduction**

Compositionally complex alloys (CCAs), including multicomponent alloys, have garnered significant attention for their distinctive mechanical and structural properties [1-10]. The broad compositional range of these alloys influences their microstructures and offers numerous opportunities to enhance both structural and mechanical performance [1-10]. Previous studies have demonstrated that CCAs exhibit behaviors distinct from conventional alloys, such as variable diffusion and complex mechanical properties, primarily due to fluctuations in the local energetic landscape [4]. These behaviors are largely attributed to local variations in atomic order and bonding, known as chemical short-range order (CSRO) induced by the composition of the alloy and the associated atom-atom nearest neighbor interactions [11]. This effect is also crucial for influencing properties like oxidation, aqueous passivation, and dislocation behavior, thus shaping the alloys' performance in various applications [12-18]. The presence of SRO strengthens the alloy by impeding dislocation motion, while its disruption promotes localized deformation[19]. This balance directly influences strain hardening and damage tolerance in CrCoNi MEAs[19]. Here, primary passivators at or above a critical component concentration react with water to form passivator-oxygen and -hydroxyl bonds which quickly approach a high surface coverage.

Recent advancements propose a method for designing anti-corrosion Cr-rich alloys by tuning their local chemical order to achieve Cr-Cr ordering in low Cr-containing CCAs and quaternary alloys [12-15]. Selected current passivation models based on the percolation of passivating element [12-13] and/or its oxide nucleation and growth [16-17] to achieve passivation are depend crucially on local arrangements of passivating elements. According to both models, there is a critical concentration of the passivating component, *P*, is needed to establish an infinite connection between *P mer* units (percolation model) or form the first $Cr_2O_3$ stable nuclei and grow via vacancy-mediated surface diffusion. However, both these models depend on the *nn* configuration of *P* such that nearer the *P* atoms (smaller bond lengths), such that active dissolution is required before high enough *Cr-O-Cr* surface coverage to stabilize passivity [12-13]. These configurations can be modified into ordered or clustered structures, such as CSRO among *P-P* atoms through specific alloy heat treatments, altering the *P-P* bond lengths [18, 20]. Studies have demonstrated that alloys with Cr-Cr ordering type CSRO delay and degrade Cr oxide/hydroxide-based passivation in Fe-Cr alloys. In contrast, clustering type CSRO is predicted to enhance it. Blades *et al.* [13] confirmed the improvement of the passivation behavior of alloys by tuning the chemical short-range order parameter. The study showed that adding only 0.03 to 0.06 mol/mol of Al to a $(FeCoNi)_{0.9}Cr_{0.1}$ alloy changed both the magnitude and sign of the Cr-Cr order parameter, resulting in passivation behavior similar to 304L stainless steel, which contains twice the amount of Cr. This approach holds promise for significantly improving the corrosion resistance of alloys, especially in severe environments.

Specifically, CSRO ordering of the primary passivating component within the first nearest neighbor atom-shell reduces the ability of atoms to find $O^{2-}$ and $OH^-$ available for passivation. This can increase the percolation threshold and delay passivation until a greater area density of primary passivators is exposed to enable passivation [12-14]. Although CSRO has been observed in CoCrNi Medium Entropy Alloys (MEAs) [11, 21-30], rigorously characterizing it to understand the atomic arrangement of nearest neighbors remains a challenge. Such characterization is essential for directly linking atomic arrangements to macroscopic properties, such as passivation behavior. A second challenge is to extend passivity to harsh environments not only those containing reducing acids but also containing $Cl^-$ which raises a question about the application of critical passivating component thresholds developed in reducing acid compared to environments with both $H^+$ and $Cl^-$.

Equiatomic CoCrNi exhibits superior corrosion resistance compared to the well-established 304 stainless steel (SS), known globally for its long-standing excellent performance [31]. Studies indicate that this alloy shows significantly enhanced performance over 304 SS in corrosive environments due to its higher concentration of alloyed Cr as the primary passivator of the three components. The passive film formed on the CoCrNi alloy is enriched in Cr but depleted in Co and Ni. The high chromium content and the thicker and more compact passive film are responsible for the CoCrNi alloy's higher corrosion resistance in acidic solution [31].

Several studies have reported that MEAs deviate from random solid solutions, exhibiting increased short-range order upon annealing [11,21-30]. For instance, Zhang *et al.* [28] used energy-filtered transmission electron microscopy (EFTEM) to observe structural features related to short-range order in CrCoNi MEAs. These findings indicate that the degree of local ordering at the nanometer scale can be adjusted through thermomechanical processing, offering a new approach to fine-tune the mechanical properties of medium- and high-entropy alloys. However, fundamental questions remain regarding the origins, chemical signatures, and stability ranges of CSRO. Another study utilizing EFTEM revealed that local chemical order increased with aging, showing a higher degree of ordering in aged samples compared to splat-quenched and 1000 °C-quenched specimens [32]. Diffuse scattering in electron diffraction patterns has been widely used to study SRO in MEAs and other concentrated FCC solid solutions. These diffuse features have traditionally been interpreted as signatures of SRO, reflecting deviations from random atomic distributions [30]. However, this interpretation has been recently questioned, with evidence suggesting that diffuse intensities can also be largely attributed to thermal vibrations and atomic displacements rather than compositional ordering[33]. One effective technique to chemically examine CSRO is extended x-ray absorption fine structure (EXAFS) measurements, which have shown promise in detecting CSRO in CoCrNi alloys. Zhang *et al.* [29] demonstrated that in the as-received CoCrNi, Cr preferentially bonds with Ni or Co, and the degree of CSRO increases following irradiation. In a related study, Smekhova *et al.* [34] used EXAFS to examine structural disorder in the CoCrNi-based alloy, revealing that the disorder is more pronounced at Cr sites than other elements. They also observed that the degree of ordering is enhanced by annealing temperature. Nevertheless, accurately fitting EXAFS data in multicomponent alloys poses substantial challenges. Recent efforts by Joress *et al.* [35] have addressed these challenges, tackling issues such as precise modeling of intricate structures, differentiation between diverse chemical environments of the same element, noise within experimental data, and the impact of theoretical assumptions on result reliability and accuracy. These efforts are crucial for advancing the use of EXAFS to understand complex multicomponent alloy systems and their CSRO-dependent properties.

Therefore, we aim to tune the local chemical order of MEA through heat treatment to enhance the alloy's corrosion resistance further. In this study, we optimize local chemical order through thermomechanical treatment to induce a greater extent of Cr-Cr clustering, to improve the corrosion resistance of MEAs. We employed EXAFS, complemented by atomic simulations, to analyze CSRO, while XPS and STEM-EELS were used for chemical characterization of the passive film. Our study demonstrates the correlation between CSRO and corrosion properties in typical CoCrNi-based alloys. The direct observation of Cr-Cr clustering illustrates enhanced short- and long-term aqueous passivation and improved performance durability in chloride-rich sulfuric acid environments. Here, chloride was added to challenge the high Cr content in solid solution in the ternary alloy such that a combination of $Cr_2O_3$ passive film formation and breakdown can occur simultaneously[36].

## Methods
### Sample Preparations and Characterization

The raw ingot of the equiatomic CoCrNi MEA was subjected to double melting in an argon-arc furnace and then sectioned into smaller pieces. These pieces were subsequently split into two groups for distinct thermal treatments: one group was as-homogenized at 1200 °C for 48 hours, followed by rapid quenching in water to room temperature; the other group received the same homogenization at 1200 °C for 48 hours, but was then aged at 1000 °C for 120 hours and allowed to cool in the furnace. Hereafter, the first condition is referred to "homogenized" and the second is "aged."

### EXAFS analysis

EXAFS spectra were collected on as-homogenized and aged CoCrNi. Photons for NIST Beamline Materials Measurements (BMM) are generated by a 3-pole wiggler. The beam was monochromated by Si (111) double crystal monochromator scanning in a pseudo-channel cut modality. The incident beam has a bandpass $\frac{\Delta E}{E}$ of around 0.01 %. The sample was illuminated using a beam size of 1 mm (vertically) by 1.5 mm (horizontally). The spectra were collected in fluorescence mode in a traditional low-Compton scattering geometry using a 4-element Vortex1 detector. The signal was normalized by an $N_{2(g)}$ ionization chamber upstream of the sample. The similarity in backscattering amplitudes makes it difficult to distinguish contributions from neighboring elements like Co, Ni in the periodic table using EXAFS. Therefore, we refer to Ni,

Co as M. The first nearest neighbor (1NN) contributions to the EXAFS signals measured at the Cr, Co, Ni absorbing atom K-edges can be expressed as:

$$\chi_i(k) = \Sigma_i \frac{S_0^2 n_i}{kR_i^2} |f_i^{eff}(k)| \sin\left[2kR_i - \frac{4}{3}k^3 + \delta_i(k)\right] \times e^{-2\sigma_i^2 k^2} e^{-2R_i/\lambda_i(k)}, \text{ Eq. (1) [37-39]}$$

where $k$ is the photoelectron wave number, $f_i^{eff}(k)$ and $\delta_i(k)$ represent the amplitude and phase of the photoelectron scattering path, respectively. $S_0^2$ is the passive electron reduction factor, $n_i$ is the degeneracy of the scattering path, $R_i$ is the effective half-path length (which equals the interatomic distance for single-scattering paths), $\sigma_i^2$ is the mean-square deviation in $R_i$, and $\lambda_i(k)$ is the photoelectron mean free path. Here, $R_i = \Delta R + R_0$, $R_0$ is the initial model interatomic distance and $\Delta R$ is the correction obtained from the fit. Detailed information about EXAFS analysis can be found in our previous work [39].

**Electrochemical Evaluations**

Electrochemical tests were conducted in deaerated (with $N_2$ gas) 0.1 mol/L $H_2SO_{4(aq)}$ and 0.1 mol/L $H_2SO_{4(aq)}$ + 3 mol/L $NaCl_{(aq)}$ using a Gamry Instrument Reference 620™ potentiostat. Solutions were prepared using reagent-grade sulfuric acid (Fisher Scientific, 98 % w/w), sodium chloride (Fisher Scientific, 99 % pure), and ultrapure water (Millipore Sigma, 18.2 MΩ·cm$^2$). Coupons of as-homogenized and aged CoCrNi were cold epoxy mounted and ground to 1200 mesh size SiC paper and polished to 1 $\mu m$ using a polycrystalline diamond suspension. Samples were cleaned via ultrasonication in acetone and ultrapure water and dried with $N_2$ gas. A standard three-electrode flat cell was used with a mercury/mercury sulfate reference electrode (E = +0.640 V vs. standard hydrogen electrode (SHE)), Pt mesh counter electrode, and alloy sample as the working electrode. All electrochemical potentials in this work are referenced to SHE. The sample exposure area was limited to 0.06 cm² by a square profile Viton O-ring. The air-formed oxide was cathodically reduced by following a three-step potentiostatic hold protocol of -0.76 V for 300 s, -1.26 V for 3 s, and -0.76 V for 60 s [12-14].

After cathodically reducing the native oxide, anodic linear sweep voltammetry (LSV) experiments were conducted from -0.3 V vs. SHE to +1.2 V vs. SHE at a scan rate of ≈ 0.52 mV/s. The imaginary impedance component ($-Z_{img}$) was monitored in-situ for every potential step of 5 mV using an applied AC sinusoidal signal of potential 20 mV (RMS) and frequency 1 Hz. Single-step chronoamperometry re-passivation experiments were conducted at +0.25 V for 600 s. Like that of LSV, the native oxide was cathodically reduced before the chronoamperometry experiments. Assuming FCC CoCrNi alloys exposing primarily the (111) plane, and that the constituent Cr, Co, and Ni metals oxidized to their +3, +2, and +2 valence states, respectively, the number of dissolved monolayers ($h$) can be calculated using the relation [12, 40]:

$$h = \int_0^t i \cdot dt \left( \frac{2 \times q \times (3 \times M_{Cr} + 2 \times (1 - M_{Cr}))}{\sqrt{3}/2 \times a^2} \right)^{-1} \quad ...Eq.(2)$$

where $i$ is the current density obtained from the chronoamperometry experiment, $t$ is the chosen period of time, that is 600 s, $q$ is the elementary unit of charge, $M_{Cr}$ is the mole fraction of Cr in the equiatomic CoCrNi, that is, and $a$ is the lattice parameter of FCC CoCrNi lattice obtained from TEM diffraction analyses. Lattice parameters of 0.3627 nm and 0.3677 nm were used for as-homogenized and aged CoCrNi MEAs, respectively, see supplementary Fig.1. Long term single step chronoamperometry experiments were performed using the following protocol: (1) cathodically reducing the air-formed oxide following the above-mentioned three-step protocol, (2) single step chronoamperometry at +0.2 V (stepped from -0.76 V) for 10 ks, and (3) a full spectrum potentiostatic electrochemical impedance spectroscopy (EIS) from 100 kHz to 1 mHz after 10 ks holds measured every 8 points per frequency decade using an AC signal of amplitude 20 mV (RMS) also at +0.2 V.

**Oxide Passive Film Compositions**

A PHI VersaProbe-III X-ray photoelectron spectroscopy (XPS) analyzer was used to obtain the cation compositions of the electrochemically grown oxide passive film after 10 ks at +0.2 V in deaerated 0.1 mol/L $H_2SO_{4(aq)}$ + 3 mol/L $NaCl_{(aq)}$. The instrument was calibrated with an Au standard to the $4f_{7/2}$ core level at 84.00 eV binding energy. C 1s (284.8 eV) was used as a reference to shift spectra. High-resolution core-shell spectra of O 1s, Co $2p_{1/2}$, Ni $2p_{3/2}$, and Cr $2p_{3/2}$ were collected using a 100 W Al $K\alpha$ X-ray source, 26 eV pass energy, and 0.05 eV time interval per step across a 100 × 100 $\mu m$ area of the sample. Spectral deconvolution peak fits were analyzed using the kolXPD software to calculate enrichment factors of Co (II), Cr (III), and Ni(II) cations, details of which can be found elsewhere [12,14, 41-42].

The passive film thickness and chemistry were further analyzed using STEM/EELS (Scanning Transmission Electron Microscopy/Electron Energy Loss Spectroscopy) with the JEOL Grand ARM2 TEM equipped with Cs-corrected STEM. TEM lamellae were prepared using the FIBSEM HELIOS G5 UC focused ion dual beam system. The samples underwent a standard FIB in-situ lift-out procedure utilizing 30 kV Ga ions and were subsequently thinned to electron transparency at 2 kV.

**Results and Discussion**

To establish a connection between local chemical order and critical observations regarding the tunable passivation threshold in CoCrNi, we examined the presence of CSRO in both the as-homogenized and aged alloys using EXAFS. This analysis was complemented by atomistic simulations performed with the Large-Scale Atomic/Molecular Massively Parallel Simulator (LAMMPS) [43] from which Warren-Cowley parameters were extracted. EXAFS has been extensively used to examine CSRO in metals and alloys and offers quantitative measurements up to the fourth shell. However, in multicomponent systems, numerous challenges can restrict the ability to obtain a good fit and accurate quantitative information [35]. Several factors complicate EXAFS analysis in CCAs, such as poor elemental contrast and the presence of interstitial oxygen [35]. The lack of distinct elemental contrast within the CoCrNi alloy complicates the EXAFS analysis, highlighting the need for meticulous approaches to overcome these challenges and advance the understanding of SRO in complex alloys. Figure 1 compares EXAFS Fourier transform magnitude (in $r$-space) for Cr, Co, and Ni absorbers in as-homogenized and aged alloys. The data shows that the first shell peaks around Cr, Co, and Ni absorbers differ between the as-homogenized and the aged alloys, indicating, according to our model, that there is a difference between the bonding environment of each atom with its respective surroundings between the two alloy conditions (Figure 1). Given the similarity of the photoelectron scattering amplitudes and phases of Co/Cr/Ni [37-39], we cannot quantify the contributions to EXAFS by the unique scattering pairs (e.g., Cr-Co, Cr-Cr, and Cr-Ni) for the as-homogenized and aged data. Alternatively, we employ a more straightforward yet stable model that approximates all nearest neighbors to the absorber as a single atomic type. We conducted a quantitative analysis using a simplified local structural model, excluding Cr scattering from the fitting of Co/Cr/Ni absorber edges. This approach analyzes the differences in the best-fit values of the effective bond length for both the as-homogenized and aged alloys (supplementary Figures 2-5).

The average effective bond lengths of Cr–Ni/Co, Ni–Ni/Co, and Co–Co/Ni bonds (Supplementary Table 1) decreased after aging. This reduction, which can result from thermal annealing or mechanical processing, may indicate a change in the type of SRO [29]. A previous study reported a decrease in the Ni–Cr bond length after irradiation and attributed it to enhanced SRO [29]. To compare our EXAFS-measured bond lengths, we also calculated the interatomic distances in pure Ni, Co, and Cr (Supplementary Table 2). Notably, the Co–Co bond length obtained from Co edge fitting in the as-homogenized alloy is slightly shorter than that in pure Co. Similarly, the Ni–Ni bond length matches that of pure Ni, and the Cr–Ni/Co bond length decreases after aging. These observations suggest that aging enhances the chemical ordering, particularly favoring Ni as the nearest neighbor around the absorber atoms. However, due to the aforementioned limitations of the EXAFS fitting in CCAs and the exclusion of Cr scattering from the fitting of Co, Ni and Cr edges, the Cr-Cr bond length could not be determined. To further understand the effect of annealing on CSRO evolution, two atomistic models were developed to compare the chemical ordering environment after (1) homogenization at 1200°C and (2) subsequent annealing treatment at 300°C. Figure 2(a-b) shows the microstructures of a single microstate of CoCrNi after equilibrating through a hybrid Monte Carlo-Molecular Dynamics (MC-MD) simulation at each temperature. The MC-MD scheme is initially applied to a randomly decorated fcc simulation cell with *a = 3.56 A*. For every 100 MC swap attempts, 10,000 MD steps were performed under the NPT ensemble. The interatomic interactions were modeled using a neural network potential for CoCrNi (Version 2) [44] that was trained via the n2p2 code [44-46]. Convergence was reached after 1 million MD timesteps, which is equivalent to 2 nanoseconds of MD time. *See the Supplementary Information Text for more details on the simulation methodology.* Local variation in chemical order is observed as a result of the MC-MD temperature during equilibration of the CoCrNi single crystal, as seen by the system-averaged Warren-Cowley Parameters depicted in **Figure 2(c)**. This figure indicates an overall enhancement in CSRO post-annealing. Notably, the interactions modeled by this potential suggest strong Ni-Ni bonding and considerable attraction between Co-Cr and Co-Co pairs. Similar to the experimental results in this study, these models also predict a preference for Cr-Cr ordering in the 2$^{nd}$ nearest neighbor shell compared to that in the 1$^{st}$ shell.

To evaluate the role of CSRO in the corrosion resistance of MEAs, the as-homogenized and aged CoCrNi samples were tested for aqueous passivation behavior in both chloride and chloride free acid environments. First, both the CoCrNi alloys compared to 304L SS (Fe-7Ni-20Cr) (at.%) show a similar $i_{crit}$, but smaller $i_{pass}$ and larger -$Z_{imag}$ in the passive region in chloride free acid underscoring the benefit of higher Cr content in CoCrNi alloy (see Supplementary Figure 7). Turning to the issue of CSRO, anodic LSV behavior in both chloride-free and chloride-containing acids after cathodically reducing the native oxides is shown in **Figure 3 (a)**. In chloride-containing acid, a significant difference was observed with aged CoCrNi (higher positive WC parameter) which showed a smaller critical ($i_{crit}$) and passive ($i_{pass}$) current density, and a larger $-Z_{imag}$ compared to the as-homogenized CoCrNi. Here, $-Z_{imag}$ offered a qualitative metric for the thickness of the passive film due to its linear dependence on the thickness of the capacitive oxide (*see Supplementary Material for more details*).[45] Thermodynamically, $Cr_2O_3$ (-2.37 eV/atom) is the most stable oxide compared to CoO (-1.14 eV/atom) and NiO (-1.22 eV/atom), and thus, $Cr_2O_3$/$Cr(OH)_3$ are expected to form at room temperature.[47] Kinetically, pure elemental Cr and Ni are also expected to passivate around -0.3 V and +0.15 V, respectively, in chloride-free 0.1 mol/L $H_2SO_4$ while pure Co actively dissolves at all anodic potentials [48-50]. In the presence of 3 mol/L NaCl, only Cr is expected to passivate at all anodic potentials, see supplementary Figure 7. This behavior also indicates that from all the atomic pairs possible in CoCrNi alloy, CSRO of Cr-Cr pairs is of most relevance to aqueous passivation as Co and Ni selectively dissolve and/or show no passivity.

Based on this information, +0.1 V and +0.2 V were chosen for further potentiostatic polarization experiments in chloride-free and chloride-containing acid environments, respectively, to further probe the effect of Cr-Cr CSRO on passivation by a $Cr_2O_3$ rich film. **Figure 3(b)** shows the average number of monolayers dissolved (*h*) in the primary passivation stage of passive film formation for both as-homogenized and aged (high WC parameter) CoCrNi alloys in chloride-free and chloride-containing acid environments. The *h* values of both the alloys were similar in chloride-free acid, consistent with previous reports for high Cr-containing alloys where any effect of CSRO on primary passivation becomes insignificant because both CoCrNi alloys contain a Cr content above the Cr threshold of 12 at. %. [36]. In chloride-added acid, the aged sample showed an *h* value of ≈ 100 monolayers compared to that of ≈280 monolayers for the as-homogenized sample, strongly suggesting an effect of the observed CSRO from aging heat treatment. In summary, the high percentage of alloyed Cr is challenged by an acidic solution of $Cl^-$ and $SO_4^{2-}$ such that differences in passivation might be discerned. Here the aged sample shows a lower h of 100 monoloayers suggesting the connection between Cr-Cr clustering and h. **Figure 3(c)** shows the monitored current decay during the hold for 10 ks in chloride-added deaerated 0.1 mol/L $H_2SO_{4(aq)}$. Once again, the as-homogenized CoCrNi alloy (i.e., low Cr-Cr WC parameter) showed poor corrosion resistance with the signature of passive film instability due to the rise in current density with time. In contrast, the current density continues to decay for the aged CoCrNi exhibiting a stable passive film in the case of Cr-Cr clustering. After 10 ks of exposure, the passive film of aged CoCrNi alloy shows a modulus of impedance, |Z|, ≈1000 times higher than that of as-homogenized CoCrNi alloy at 1 mHz (see **Figure 3(d)**). This impedance represents ability of the Cr(III) rich passive film to regulate metal oxidation and dissolution. **Figure 3(e)** reports the XPS-based passive film cationic enrichment factors of passive films after 10 ks of single-step passivation experiment at +0.2 V in chloride added deaerated 0.1 mol/L $H_2SO_{4(aq)}$. The results suggest a slightly higher enrichment of Cr(III) cations in the passive film of aged CoCrNi (i.e., high Cr-Cr WC parameter) compared to that of as-homogenized CoCrNi with a lower WC parameter. The individual cation fractions can be found in the Supplementary material.

The short and long-term passivation difference with WC parameter can be understood from a simultaneous occurrence of $Cr_2O_3$ passive film formation, and re-passivation in the presence of $Cl^-$ ions, as shown in **Figure 4**.[12,36,50] Following the passivation models, during passive film formation (primary passivation), Cr atoms bond with dissolved oxygen (from adsorbed $OH^-$) to form a network of *Cr–O–Cr mer* units [47] or form nuclei of $Cr_2O_3$. Once these *mer* units or nuclei are either infinitely connected via percolation or surface diffusion of Cr atoms, respectively, the surface chain evolves into a monolayer (surface coverage) of the oxide passive film consisting of $Cr(OH)_3/Cr_2O_3$. The active $Cl^-$ anions are subject to competitive adsorption with $O^{2-}$ and $OH^-$, where the adsorbed $Cl^-$ is expected to bond with Cr atoms and form $CrCl_3$, competing with the formation of *Cr–O–Cr* bonds. This is reflected in a more active dissolution of the alloy in terms of higher $h$ and $i_{crit}$ of the alloy needed for passive film formation when the Cr-Cr WC parameter has a less negative value (see **Figure 2**). The passive film chemistry and thickness of the oxide scale using STEM/EELS, shown in **Figure 5,** underscore what is indicated by the corrosion electrochemistry. XPS reveals that the oxide thickness in the aged alloy is greater compared to that of the as-homogenized alloy. Specifically, the average oxide passive film thickness increased from approximately 7 nm for the as-homogenized alloy to ≈ 11 nm for the aged sample. This suggests that passive film formation and ensuing protection are greater for the aged sample [45-53]. The chemical analysis revealed that Cr (III) oxide is more present to a greater extent in the passive film in the aged alloy compared to the as-homogenized alloy.

Changes in alloy microstructure induced by the thermal aging treatment can also reduce microstructural defects, such as dislocations, grain boundary density, grain boundary triple points, or grain size [36,54], making it less susceptible to pitting and/or facilitating passivation. The aged sample was found to have a slightly greater grain size of 85.6 μm than the homogenized sample of 60.9 μm, see Supplementary Figure 8. This should produce fewer grain boundary pit sites and possibly induce a small decrease in pitting susceptibility that would be detected in the impedance data for aged vs. homogenized. The impedance data should be lower in the homogenized condition, but should also become very noisy, as well as the drop from the large values where pitting occurs, features that could account for the changes observed. This is not evident in impedance data as shown in **Figure 3**. Moreover, $Z_{imag}$ should also be disrupted. It should be noted that this in addition to the alteration of CSRO due to aging, these effects on pitting were not observed.

**Summary**

Given that the passive films of both the as-homogenized and aged alloys were grown under identical electrochemical conditions, the differences in short- and long-term corrosion behavior, as well as film thickness, can be attributed to variations in CSRO caused by the aging treatment, which influences the passivation processes. Moreover, the Cr-Cr clustering results from an aging process, indicating a general approach of introducing CSRO in alloys containing three or more elements. This suggests a design strategy for enhancing passivation behavior by promoting clustering of the passivating element through thermal heat treatments.

**Acknowledgments**

This work was performed (in part) at the Materials Characterization and Processing Center (MCP) in the Whiting School of Engineering at Johns Hopkins University. The authors acknowledge partial funding, in part (E.A.A. and M.L.T.), from the National Science Foundation MRI program (Award No. 1429661) and, in part, from the Office of Naval Research through the Multidisciplinary University Research Initiative (MURI) program (Award No. N00014-20-1-2368) with program manager Dr. D. Shifler. D.S. acknowledges the partial support from the UVA-DMSE Olsen Graduate Fellowship during the period of this work. A. I. F. acknowledges support from the National Science Foundation (Grant No. CHE 2203858) for his contribution to the EXAFS data analysis and interpretation. Additionally, the authors acknowledge funding from the United States Department of Energy, Office of Basic Energy Sciences through contract DE-SC0020314. The authors acknowledge the University of Virginia Nanoscale Materials Characterization Facility (NMCF) for utilizing the PHI Versaprobe III XPS that was supported by NSF Award #162601. This research used resources at the 6-BMM beamline of the National Synchrotron Light Source II, a U.S. Department of Energy (DOE) Office of Science User Facility operated for the DOE Office of Science by Brookhaven National Laboratory under Contract No. DE-SC0012704. AM was supported by the US Department of Energy, Office of Science, Basic Energy Sciences, Materials Sciences and Engineering Division, through the Damage-Tolerance in Structural Materials program (KC13) at the Lawrence Berkeley National Laboratory (LBNL) under contract DE-AC02-CH11231. The authors are grateful to Dr. Bruce Ravel (National Institute of Standards and Technology), for assistance and guidance for successful EXAFS measurements and Ruopeng Zhang (UC Berkeley) for the jet-polished NiCoCr samples. Certain equipment, instruments, software, or materials are identified in this paper in order to specify the experimental procedure adequately. Such identification is not intended to imply recommendation or endorsement of any product or service, nor is it intended to imply that the materials or equipment identified are necessarily the best available for the purpose.

**References**

1. Miracle, D. B., & Senkov, O. N. (2017). A critical review of high entropy alloys and related concepts. Acta Materialia, 122, 556–511. https://doi.org/10.1016/j.actamat.2016.08.081

2. Yeh, J. W., Chen, S., Lin, S., Gan, J., Chin, T., Shun, T., & Tsau, C. (2004). Nanostructured high-entropy alloys with multiple principal elements: Novel alloy design concepts and outcomes. Advanced Engineering Materials, 6(5), 299–303. https://doi.org/10.1002/adem.200300567

3. Cantor, B., Chang, I. T. H., Knight, P., & Vincent, A. J. B. (2004). Microstructural development in equiatomic multicomponent alloys. Materials Science and Engineering: A, 375–377, 213–218. https://doi.org/10.1016/j.msea.2003.10.257

4. Zhang, Y., Zuo, T. T., Tang, Z., Gao, M. C., Dahmen, K. A., Liaw, P. K., & Lu, Z. P. (2014). Microstructures and properties of high-entropy alloys. Progress in Materials Science, 61, 1–93. https://doi.org/10.1016/j.pmatsci.2013.10.001

5. Senkov, O. N., Wilks, G. B., Scott, J. M., & Miracle, D. B. (2011). Mechanical properties of $Nb_{25}Mo_{25}Ta_{25}W_{25}$ and $V_{20}Nb_{20}Mo_{20}Ta_{20}W_{20}$ high entropy alloys. Intermetallics, 19, 689–693. https://doi.org/10.1016/j.intermet.2011.01.017

6. Otto, F., Dlouhy, A., Bei, H., Ridge, O., & Eggeler, G. (2013). The influences of temperature and microstructure on the tensile properties of a CoCrFeMnNi high-entropy alloy. Acta Materialia, 61(16), 6745–6757. https://doi.org/10.1016/j.actamat.2013.06.018

7. Gwalani, B., Gorsse, S., Choudhuri, D., Styles, M., Zheng, Y., Mishra, R. S., & Banerjee, R. (2018). Modifying transformation pathways in high entropy alloys or complex concentrated alloys via thermo-mechanical processing. Acta Materialia, 153, 169–185. https://doi.org/10.1016/j.actamat.2018.05.037

8. Beke, D. L. (2016). Stability and metastability of phases in nanosized systems. Materials Letters, 164, 472–475.

9. Zhang, Y., Stocks, G. M., Jin, K., Lu, C., Bei, H., Sales, B. C., Weber, W. J., & Zhang, Y. (2015). Influence of chemical disorder on energy dissipation and defect evolution in concentrated solid solution alloys. Nature Communications, 6, 8736. https://doi.org/10.1038/ncomms9736

10. Barr, C. M., LaRosa, C. R., & Thompson, G. B. (2020). Mechanisms for phase separation in high entropy alloys. Scripta Materialia, 176, 6–10. https://doi.org/10.1016/j.scriptamat.2019.09.035

11. Zhang, Q., Jin, X., Shi, X. H., Qiao, J. W., & Liaw, P. K. (2022). Short-range ordering and strengthening in CoCrNi medium-entropy alloy. Materials Science and Engineering: A, 854, 143890. https://doi.org/10.1016/j.msea.2022.143890

12. Xie, Y., Artymowicz, D. M., Lopes, P. P., & Sieradzki, K. (2021). A percolation theory for designing corrosion-resistant alloys. Nature Materials, 20(6), 789–793. https://doi.org/10.1038/s41563-021-00920-0

13. W. Blades, B. Redemann, N. Smith, et al., "Tuning chemical short-range order for stainless behavior at reduced chromium concentrations in multi-principal element alloys," Acta Materialia, vol. 277, p. 120209, Sep. 2024. DOI: 10.1016/j.actamat.2024.120209.

14. D. Sur, N. Smith, P. F. Connors, W. H. Blades, M. L. Taheri, C. M. Wolverton, K. Sieradzki, and J. R. Scully, Investigating the long-term synergistic benefits of al on Cr(III) in the passive films of FeCoNi-Cr-Al CCAs in sulfuric acid, 2024, doi.org/10.2139/ssrn.4984641

15. D. Sur, S. B. Inman, K. L. Anderson, N. Smith, J. Qi, C. W. Wolverton, J. R. Scully, "Factors governing passivation behavior of Fe-Cr-Al-Ti alloys in sulfate containing acidic solutions: uncovering the many roles of Ti", Materialia, 39, 102370. 2025. https://doi.org/10.1016/j.mtla.2025.102370

16. Diawara, B., Beh, Y. A., & Marcus, P. (2010). Nucleation and growth of oxide layers on stainless steels (FeCr) using a virtual oxide layer model. *Journal of Physical Chemistry C*, *114*(45), 19299–19307. https://doi.org/10.1021/jp909445x

17. Legrand, M., Diawara, B., Legendre, J.-J., & Marcus, P. (2002). Three-dimensional modelling of selective dissolution and passivation of iron–chromium alloys. *Corrosion Science*, *44*(4), 773–790. https://doi.org/10.1016/S0010-938X(01)00073-7

18. Schweika, W., & Haubold, H.-G. (1988). Neutron-scattering and Monte Carlo study of short-range order and atomic interaction in Ni0.89Cr0.11. *Physical Review B*, *37*(16), 9240–9248. https://doi.org/10.1103/PhysRevB.37.9240

19. Yang, Y., Yin, S., Yu, Q., Zhu, Y., Ding, J., Zhang, R., Ophus, C., Asta, M., Ritchie, R. O., & Minor, A. M. (2024). Rejuvenation as the origin of planar defects in the CrCoNi medium entropy alloy. *Nature Communications, 15*, Article 1402. https://doi.org/10.1038/s41467-024-42650-6

20. Smekhova, A., Kuzmin, A., Siemensmeyer, K., Luo, C., Taylor, J., Thakur, S., Radu, F., Weschke, E., Buzanich, A. G., Xiao, B., Savan, A., Yusenko, K. V., & Ludwig, A. (2022). Structural disorder and magnetic properties of a nanocrystalline Cantor alloy film down to the atomic scale. Spectroscopy Europe. Available at: https://www.spectroscopyeurope.com/news/exafs-reveals-structural-disorder-and-magnetic-properties-high-entropy-alloys

21. Gludovatz, B., Hohenwarter, A., Thurston, K. V. S., Bei, H., Wu, Z., & George, E. P. (2016). Exceptional damage-tolerance of a medium-entropy alloy CrCoNi at cryogenic temperatures. Nature Communications, 7, 10602. https://doi.org/10.1038/ncomms10602

22. Liu, Y., Zhang, H., Yang, Y., Sun, L., Zhao, X., Yan, H.-L., Shen, Y., & Jia, N. (2023). Chemical short-range order dependence of micromechanical behavior in CoCrNi medium-entropy alloy studied by atomic simulations. Journal of Alloys and Compounds, 968, 172002. https://doi.org/10.1016/j.jallcom.2023.172002

23. Jian, W.-R., Xie, Z., Xu, S., Su, Y., Yao, X., & Beyerlein, I. J. (2020). Effects of lattice distortion and chemical short-range order on the mechanisms of deformation in medium entropy alloy CoCrNi. Acta Materialia, 199, 352–369. https://doi.org/10.1016/j.actamat.2020.06.041

24. Yang, X., Xi, Y., He, C., Chen, H., Zhang, X., & Tu, S. (2022). Chemical short-range order strengthening mechanism in CoCrNi medium-entropy alloy under nanoindentation. Scripta Materialia, 209, 114364. https://doi.org/10.1016/j.scriptamat.2022.114364

25. Xu, D., Wang, M., Li, T., Wei, X., & Lu, Y. (2022). A critical review of the mechanical properties of CoCrNi-based medium-entropy alloys. Microstructures, 2(1), 2022001. https://doi.org/10.20517/microstructures.2021.10

26. Ziehl, T. J., Morris, D., & Zhang, P. (2023). Detection and impact of short-range order in medium/high-entropy alloys. iScience, 26, 106209. https://doi.org/10.1016/j.isci.2023.106209

27. Ferrari, A., Körmann, F., Asta, M., & Neugebauer, J. (2023). Simulating short-range order in compositionally complex materials. Nature Computational Science, 3(3), 221–229. https://doi.org/10.1038/s43588-023-00407-4

28. Zhang, R., Zhao, S., Ding, J., et al. (2020). Short-range order and its impact on the CrCoNi medium-entropy alloy. Nature, 581(7808), 283–287. https://doi.org/10.1038/s41586-020-2275-z

29. Zhang, F. X., Zhao, S., Jin, K., Xue, H., Velisa, G., Bei, H., Huang, R., Ko, J. Y. P., Pagan, D. C., Neuefeind, J. C., Weber, W. J., & Zhang, Y. (2017). Local structure and short-range order in a NiCoCr solid solution alloy. Physical Review Letters, 118(20), 205501. https://doi.org/10.1103/PhysRevLett.118.205501

30. Foley, D. L., Barnett, A. K., Rakita, Y., Perez, A., Das, P. P., Nicolopoulos, S., Spearot, D. E., Beyerlein, I. J., Falk, M. L., & Taheri, M. L. (2024). Diffuse electron scattering reveals kinetic frustration as origin of order in CoCrNi medium entropy alloy. Acta Materialia, 268, 119753. https://doi.org/10.1016/j.actamat.2024.119753

31. Wang, J., Li, W., Yang, H., Huang, H., Ji, S., Ruan, J., & Liu, Z. (2020). Corrosion behavior of CoCrNi medium-entropy alloy compared with 304 stainless steel in H2SO4 and NaOH solutions. Corrosion Science, 177, 108973. https://doi.org/10.1016/j.corsci.2020.108973

32. Tsai, H.-C., Chou, Y.-C., Lin, J.-Y., & Tu, K.-T. (2022). Determination of peak ordering in the CrCoNi medium-entropy alloy via nanoindentation. *Acta Materialia, 241*, 118380. https://doi.org/10.1016/j.actamat.2022.118380

33. Flynn Walsh, M., Zhang, M., Ritchie, R. O., Asta, M., & Minor, A. M. (2024). Multiple origins of extra electron diffractions in fcc metals. *Science Advances, 10*(31), eadn9673. https://doi.org/10.1126/sciadv.adn9673

34. Smekhova, A., Kuzmin, A., Siemensmeyer, K., Luo, C., Taylor, J., Thakur, S., Radu, F., Weschke, E., Buzanich, A. G., Xiao, B., Savan, A., Yusenko, K. V., & Ludwig, A. (2022). Structural disorder and magnetic properties of a nanocrystalline Cantor alloy film down to the atomic scale. Spectroscopy Europe. Available at:
21

https://www.spectroscopyeurope.com/news/exafs-reveals-structural-disorder-and-magnetic-properties-high-entropy-alloys

35. Joress, H., Ravel, B., Anber, E., Hattrick-Simpers, J., Taheri, M. L., & DeCost, B. (2023). Why is EXAFS for complex concentrated alloys so hard? Challenges and opportunities for measuring ordering with X-ray absorption spectroscopy. Matter. https://doi.org/10.1016/j.matt.2023.09.010

36. Marcus, P., Maurice, V., & Strehblow, H.-H. (2008). Localized corrosion (pitting): A model of passivity breakdown including the role of the oxide layer nanostructure. *Corrosion Science, 50*(9), 2698–2704. https://doi.org/10.1016/j.corsci.2008.06.047

37. Frenkel, A. I. (2012). Applications of extended X-ray absorption fine-structure spectroscopy to studies of bimetallic nanoparticle catalysts. Chemical Society Reviews, 41(24), 8163–8178. https://doi.org/10.1039/C2CS35142B

38. Frenkel, A. I., Machavariani, V. Sh., Rubshtein, A., Rosenberg, Y., Voronel, A., & Stern, E. A. (2000). Local structure of disordered Au-Cu and Au-Ag alloys. Physical Review B, 62(14), 9364–9370. https://doi.org/10.1103/PhysRevB.62.9364

39. Anber, E. A., Foley, D. L., Hart, J. L., Joress, H., DeCost, B., Doherty, R., Liaw, P. K., Farkas, D., Frenkel, A. I., & Taheri, M. L. (2025). Influence of short-range order on precipitate orientation relationships in aluminum containing FCC high entropy alloys. *Intermetallics, 184*, 108832. https://doi.org/10.1016/j.intermet.2025.108832

40. Xie, S., Sieradzki, K., Yang, Y., & Zhang, L. (2016). Chemical short-range order and its effect on the mechanical properties of CoCrFeNi and CoCrFeMnNi high-entropy alloys. *Nature Materials, 15*(5), 556–563. https://doi.org/10.1038/nmat4587

41. S. Inman, D. Sur, J. Han, K. M. Ogle, and J. R. Scully, "Corrosion Behavior of a Compositionally Complex Alloy Utilizing Simultaneous Al, Cr, and Ti Passivation," Corrosion Science, vol. 217, pp. 111–138, Mar. 2023. DOI: 10.1016/j.corsci.2023.111138

42. Sur, D., Holcombe, E. F., Blades, W. H., et al. (2023). An experimental high-throughput to high-fidelity study towards discovering Al–Cr containing corrosion-resistant compositionally complex alloys. High Entropy Alloys & Materials, 1, 336–353. https://doi.org/10.1007/s44210-023-00020-0

43. Thompson, A. P., Aktulga, H. M., Berger, R., Bolintineanu, D. S., Brown, W. M., Crozier, P. S., in't Veld, P. J., Kohlmeyer, A., Moore, S. G., Nguyen, T. D., & LAMMPS Development Team. (2022). LAMMPS: A flexible simulation tool for particle-based materials modeling at the atomic, meso, and continuum scales. Computer Physics Communications, 271, 108171. https://doi.org/10.1016/j.cpc.2022.108171

44. Du, J. P., Yu, P., Shinzato, S., Meng, F. S., Sato, Y., Li, Y., Fan, Y., & Ogata, S. (2022). Chemical domain structure and its formation kinetics in CrCoNi medium-entropy alloy. *Acta Materialia, 240*.

45. B.Hirschorn, M.E.Orazem, B.Tribollet, V. Vivier, I. Frateur, and M. Musiani, "Determination of effective capacitance and film thickness from constant-phase-element

parameters," Electrochimica Acta, vol. 55, no. 21, pp. 6218–6227, Aug. 2010. DOI: 10.1016/j. electacta.2009.10.065.

46. Singraber, A.; Behler, J.; Dellago, C. Library-Based LAMMPS Implementation of High-Dimensional Neural Network Potentials. J. Chem. Theory Comput. 2019 15 (3), 1827–1840.
47. A. Jain, S. P. Ong, G. Hautier, *et al.*, "Commentary: The Materials Project: A materials genome approach to accelerating materials innovation," *APL Materials*, vol. 1, no. 1, p. 011002, Jul. 2013. DOI: 10.1063/1.4812323.
48. Lutton, K., Han, J., Ha, H. M., Sur, D., Romanovskaia, E., & Scully, J. R. (2023). Passivation of Ni-Cr and Ni-Cr-Mo Alloys in Low and High pH Sulfate Solutions. *Journal of Electrochemical Society*, *170*(2), 021507. https://doi.org/10.1149/1945-7111/acb9c3.
49. Li, X., Zhou, P., Feng, H., Jiang, Z., Li, H., & Ogle, K. (2022). Spontaneous passivation of the CoCrFeMnNi high entropy alloy in sulfuric acid solution: The effects of alloyed nitrogen and dissolved oxygen. *Corrosion Science*, *196*(July 2021), 110016. https://doi.org/10.1016/j.corsci.2021.110016
50. K. O. Sarfo, P. Murkute, O. B. Isgor, Y. Zhang, J. Tucker, and L. Arnad´ottir, "Density Func-´ tional Theory Study of the Initial Stages of Cl-Induced Degradation of α-Cr2O3 Passive Film," *Journal of The Electrochemical Society*, vol. 167, no. 12, p. 121508, Sep. 2020. DOI:10.1149/1945-7111/ABB381.1988
51. J. R. Scully, "Polarization resistance method for determination of instantaneous corrosion rates," *Corrosion*, vol. 56, no. 2, pp. 199–217, 2000. DOI: 10.5006/1.3280536
52. B. Hirschorn, M. E. Orazem, B. Tribollet, V. Vivier, I. Frateur, and M. Musiani, "ConstantPhase-Element Behavior Caused by Resistivity Distributions in Films: I. Theory," *Journal of The Electrochemical Society*, vol. 157, no. 12, p. C452, 2010. DOI: 10.1149/1.3499564.
53. B. Hirschorn, M. E. Orazem, B. Tribollet, V. Vivier, I. Frateur, and M. Musiani, "ConstantPhase-Element Behavior Caused by Resistivity Distributions in Films. II. Applications," *Journal of The Electrochemical Society*, vol. 157, no. 12, p. C458, 2010. DOI: 10.1149/1.3499565.
54. Barr, C.M., Thomas, S., Hart, J.L. *et al.* Tracking the evolution of intergranular corrosion through twin-related domains in grain boundary networks. *npj Mater Degrad* **2**, 14 (2018). https://doi.org/10.1038/s41529-018-0032-7

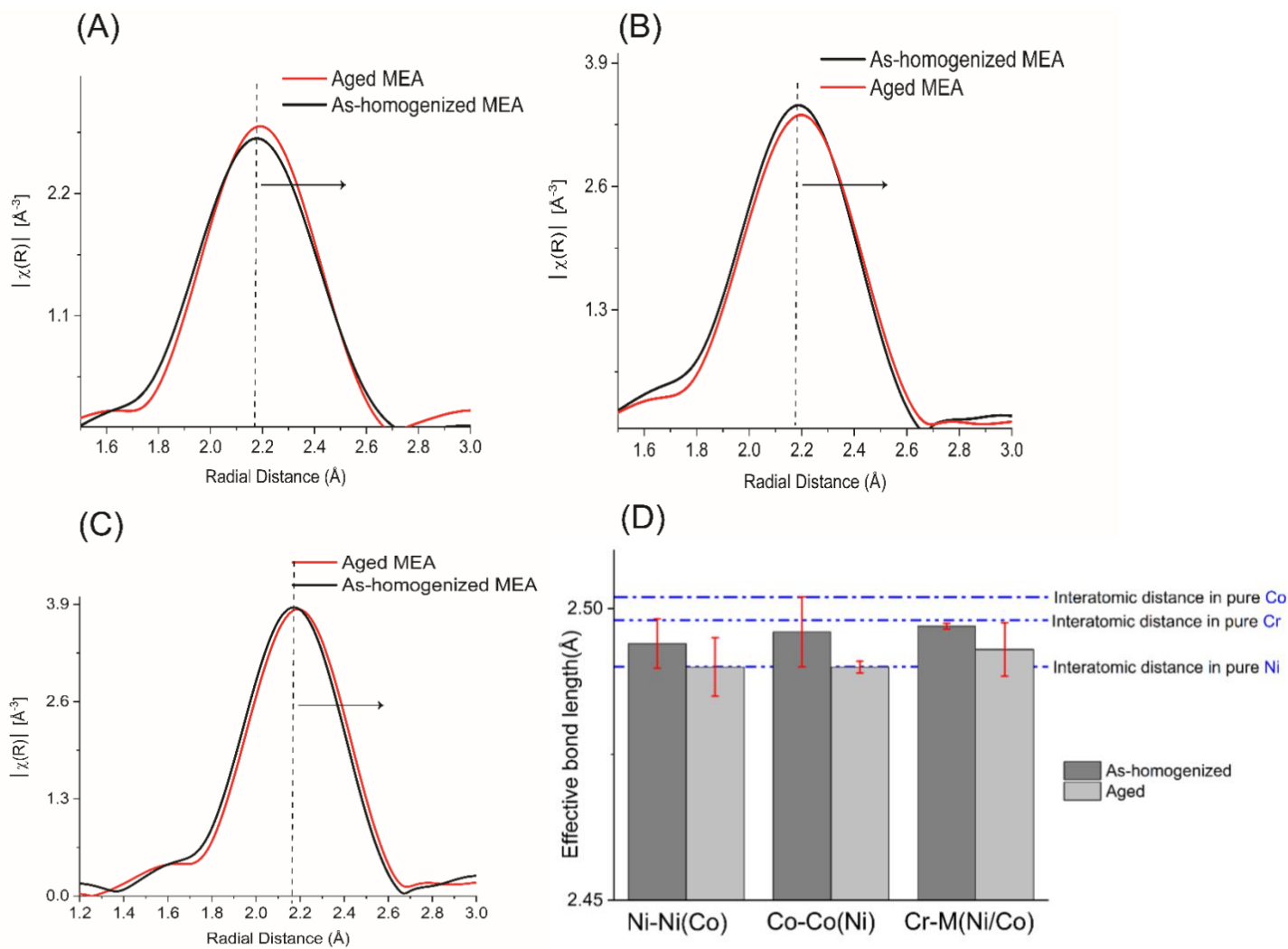

**Fig. 1**. The extracted extended edge spectra of Cr, Co, and Ni K using EXAFS/EXELFS in R space of as as-homogenized CoCrNi(A-C). The change in bond length(D) after aging may indicate that the bonding environment of Cr, Co, and Ni are different relative to their environment. The error bars represent the uncertainties in the changes in bond length magnitudes (ΔR). The blue dashed line in Fig. D represents the calculated interatomic distances in pure Ni, Co, and Cr; more details can be found in Supplementary Table 2.

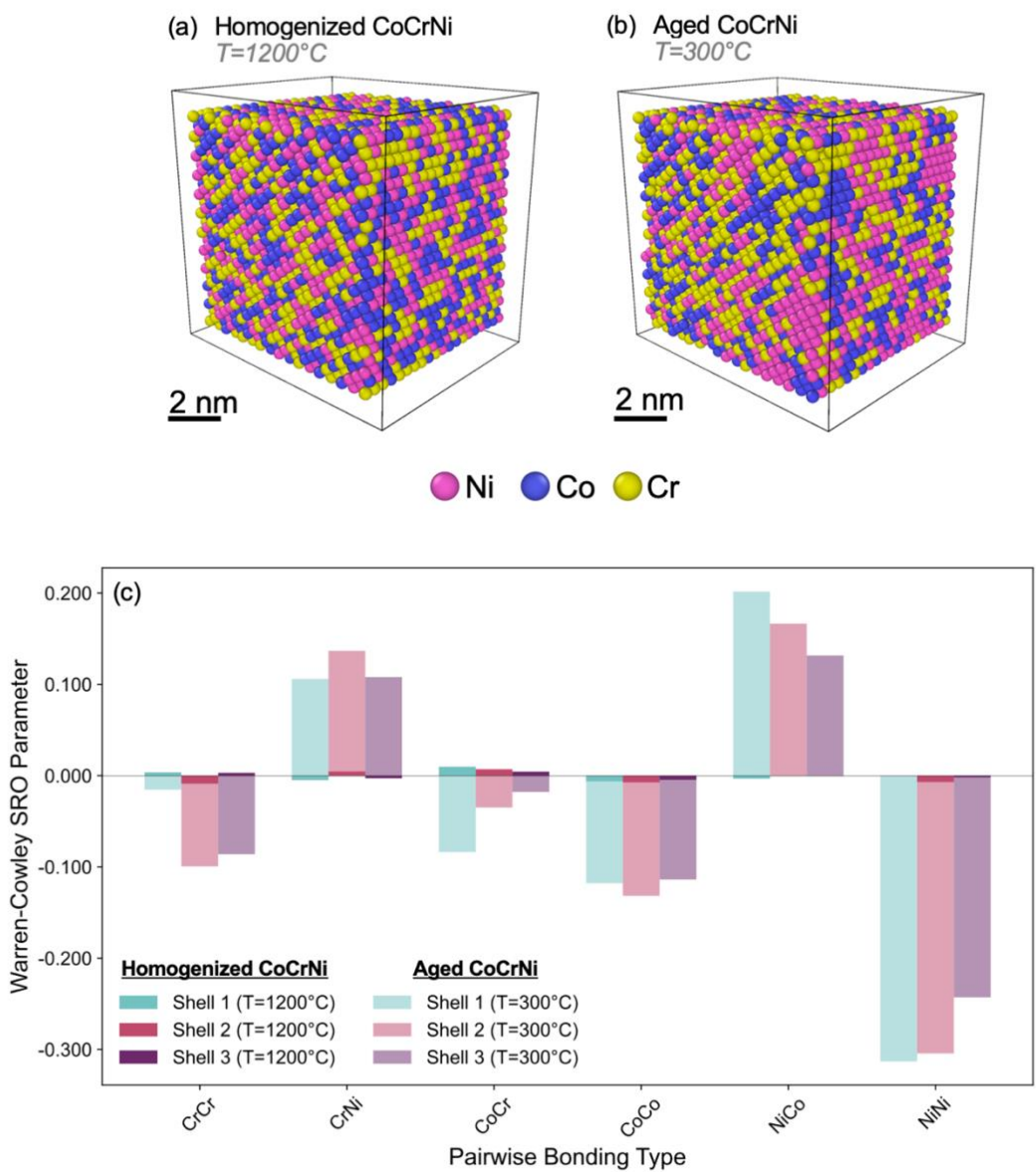

**Fig. 2.** Simulated Warren Cowley short range order (SRO) parameters indicate an increase in chemical ordering after aging at low temperatures through MCMD equilibration. Attractive bonds for both like and dislike pairs are denoted using a negative WCP value. Atomistic models after a (a) high temperature homogenization and (b) aging at 300°C show variation in (c) Warren-Cowley SRO parameters. Most notably, there is strong 2$^{nd}$ shell Cr-Cr ordering, in addition to attraction between Ni-Ni, Co-Cr, and Co-Co pairs in the low temperature model.

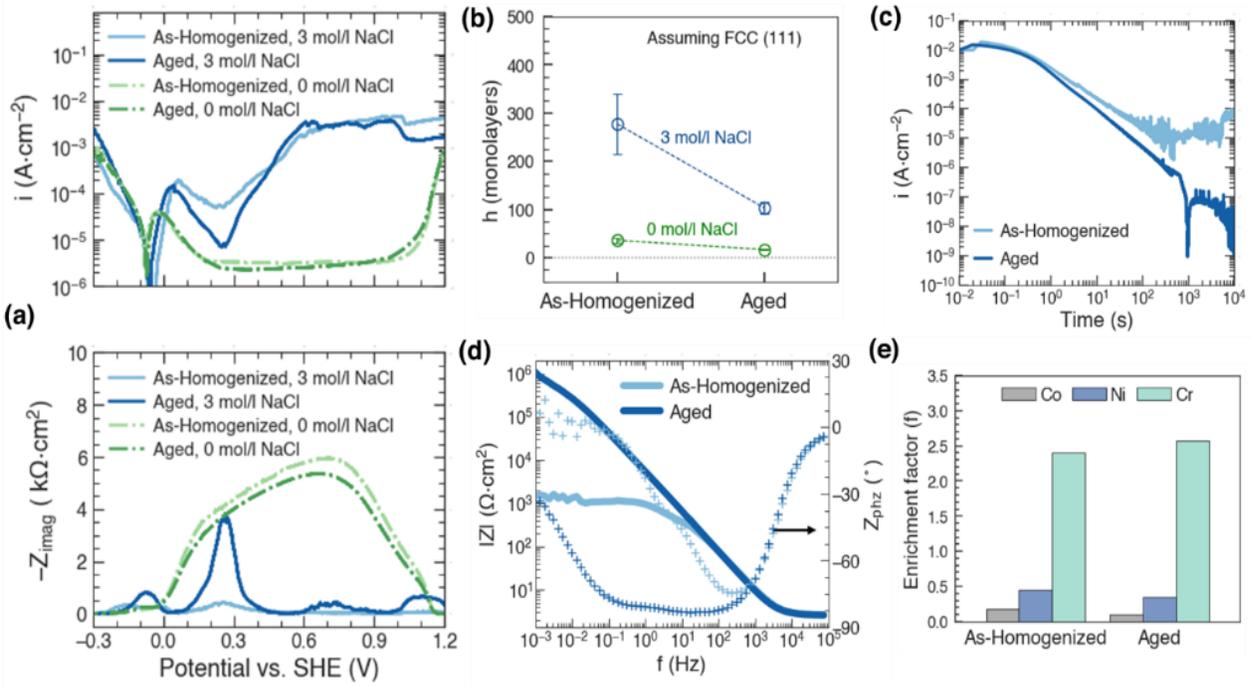

**Fig.3.** Electrochemical and oxide passive film characterization results of as-homogenized and aged CoCrNi alloys. (a) Anodic single frequency polarization curves of current density (*i*) and $-Z_{imag}$ vs. applied potential, scan rate ≈ 0.5 mV/s, (b) Calculated no. of dissolved monolayers (*h*) in both deaerated 0.1 mol/L $H_2SO_4$ and deaerated 3 mol/l NaCl added 0.1 mol/L $H_2SO_4$. (c) Current decay during the potentiostatic hold at +0.2 V and (d) EIS Bode impedance magnitude (|Z|) and phase ($Z_{phz}$) plots, and (e) XPS based passive film enrichment factors (*f*) of Co(II), Cr(III), and Ni(II) cations after 10 ks potentiostatic hold at +0.2 V in deaerated 3 mol/l NaCl added 0.1 mol/L $H_2SO_4$.

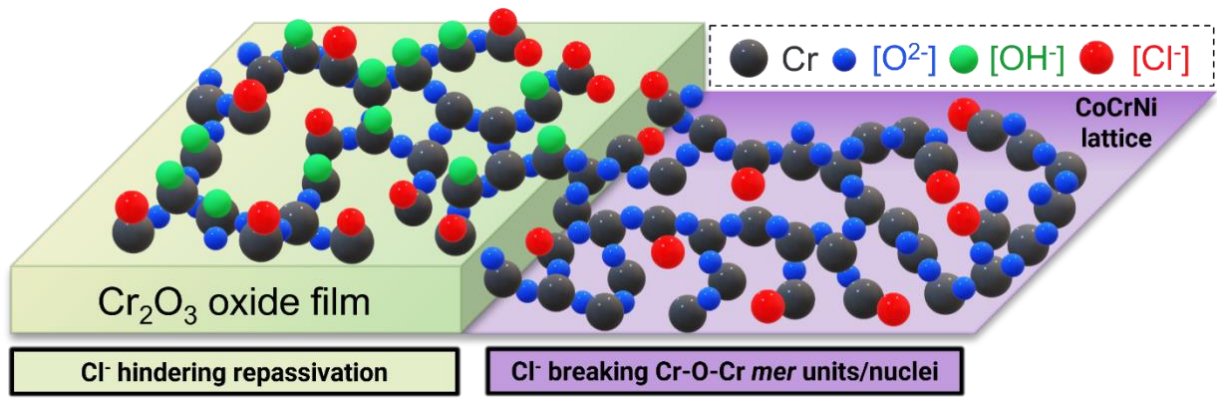

**Fig 4.** A schematic describing the passivation and re-passivation processes following the competitive adsorption between hydroxyl and chloride anions for a $Cr_2O_3/Cr(OH)_3$ passive film. **[12, 17].**

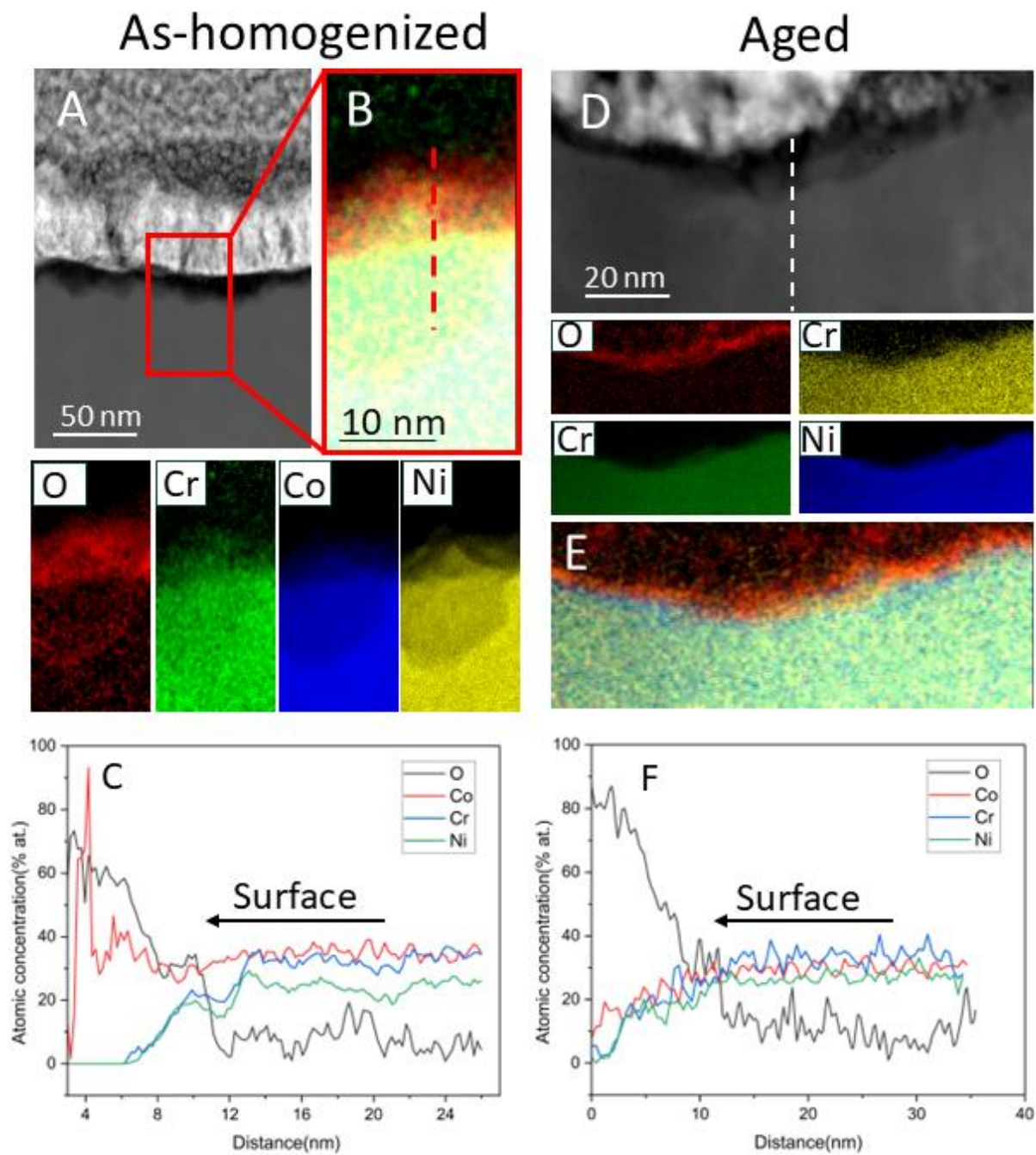

Fig. 5. Cross-sectional STEM analysis of the corroded MEA, showing the as-homogenized alloy (left) and the aged alloy (right), along with corresponding chemical mapping and line profiles (C and F) obtained using EELS. The results indicate that the passivation scale is thicker with more chromium content after aging.

# Supplemental Information

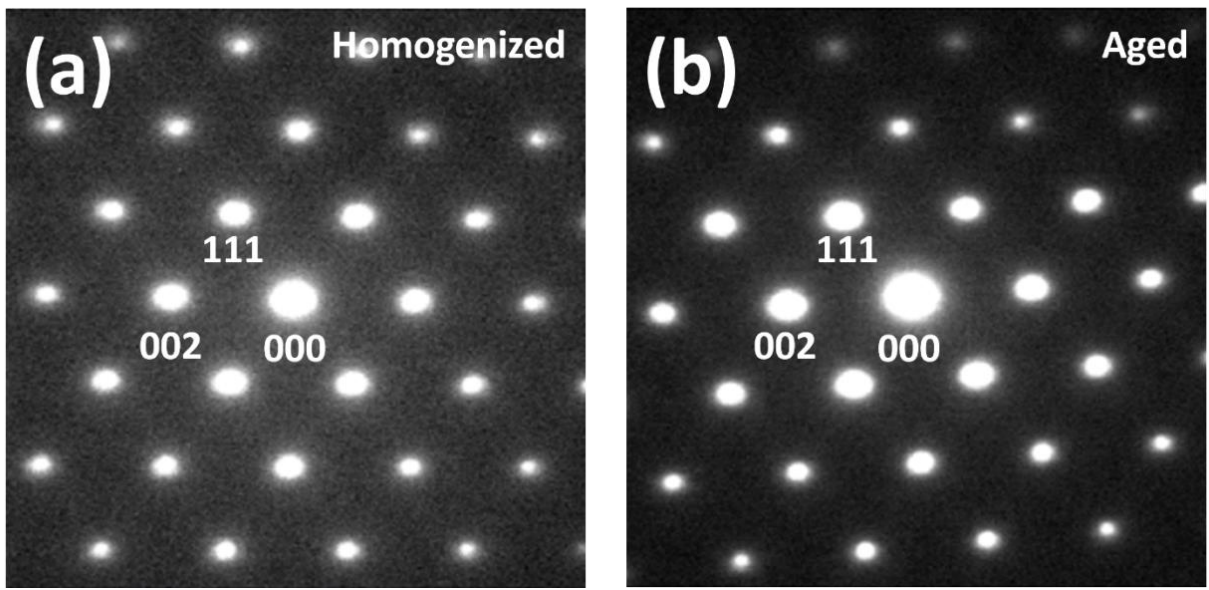

**Supplementary Fig 1.** Electron diffraction pattern along the [011] zone axis from the as-homogenized (a) and aged (b) CoCrNi alloy, showing reflections from the (111) and (002) planes.

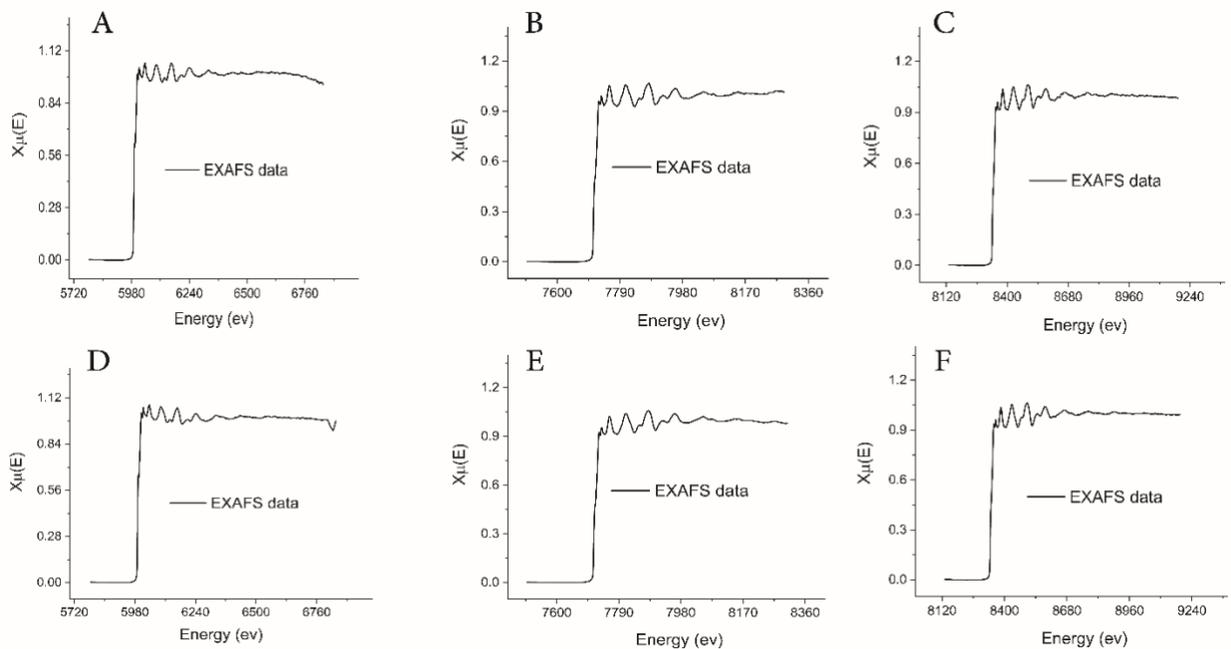

**Supplementary Fig 2**. X-ray absorption coefficients of the as-homogenized CoCrNi alloy for (a) Cr, (b) Co, and (c) Ni K-edges. The coefficients for the aged CoCrNi alloy are presented for (d) Cr, (e) Co, and (f) Ni.

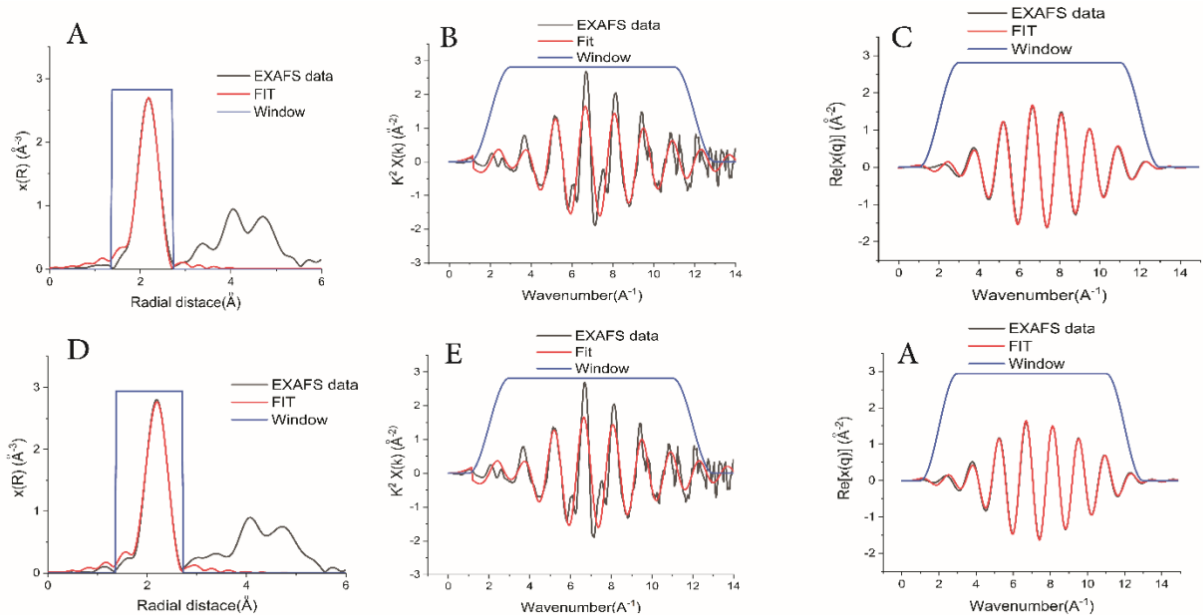

**Supplementary Fig 3.** Fourier-filtered EXAFS spectra at the Cr K-edge in R-space for the (A) as-homogenized and (D) aged alloys. EXAFS spectra at the Cr K-edge in k-space for the (B) as-homogenized and (E) aged alloys. Extracted EXAFS spectra at the Cr K-edge transformed into q-space for the (C) as-homogenized and (F) aged alloys.

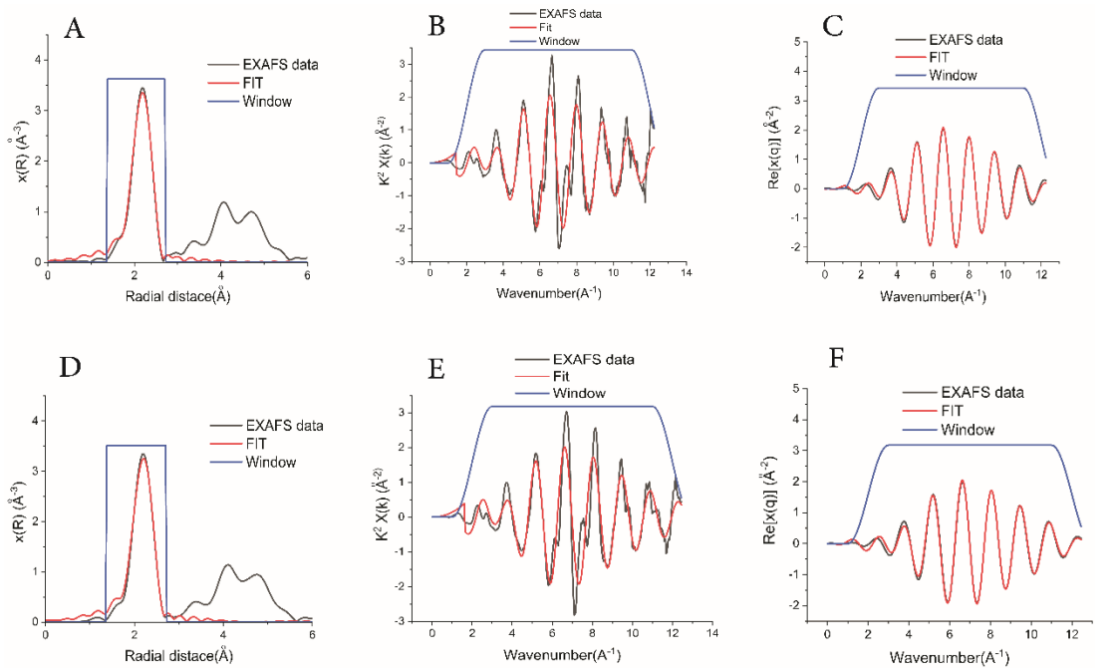

**Supplementary Fig 4.** Fourier-filtered EXAFS spectra at the Co K-edge in R-space for the (A) as-homogenized and (D) aged alloys. EXAFS spectra at the Cr K-edge in k-space for the (B) as-homogenized and (E) aged alloys. Extracted EXAFS spectra at the Cr K-edge transformed into q-space for the (C) as-homogenized and (F) aged alloys.

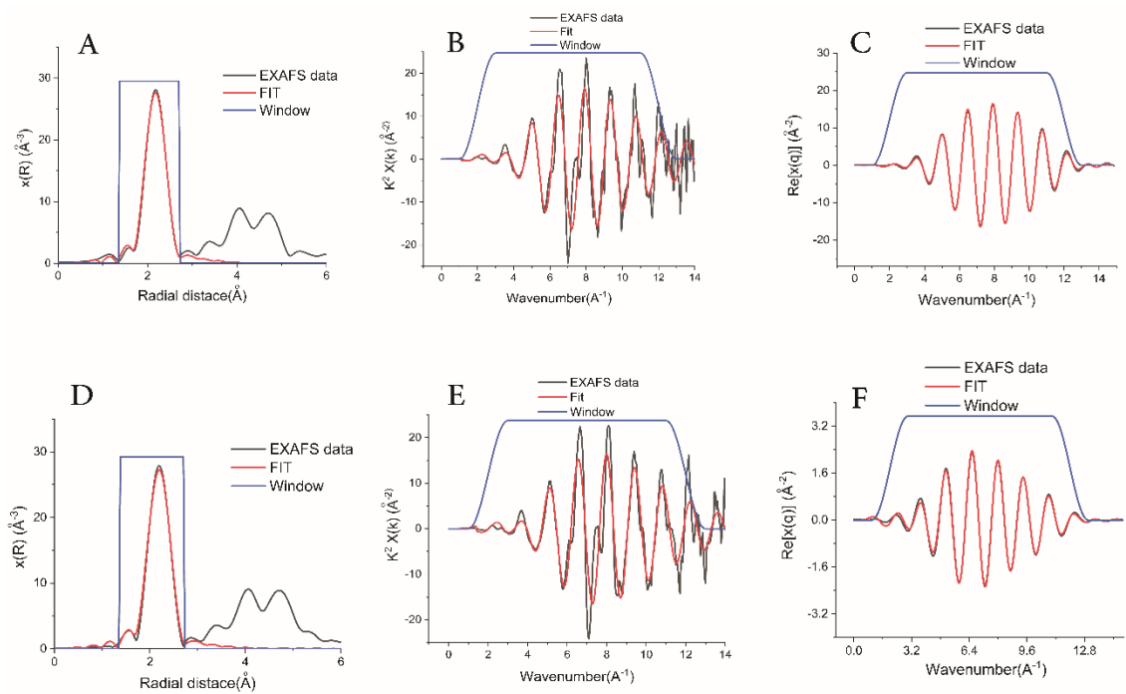

**Supplementary Fig 5**. Fourier-filtered EXAFS spectra at the Ni K-edge in R-space for the (A) as-homogenized and (D) aged alloys. EXAFS spectra at the Cr K-edge in k-space for the (B) as-homogenized and (E) aged alloys. Extracted EXAFS spectra at the Cr K-edge transformed into q-space for the (C) as-homogenized and (F) aged alloys.

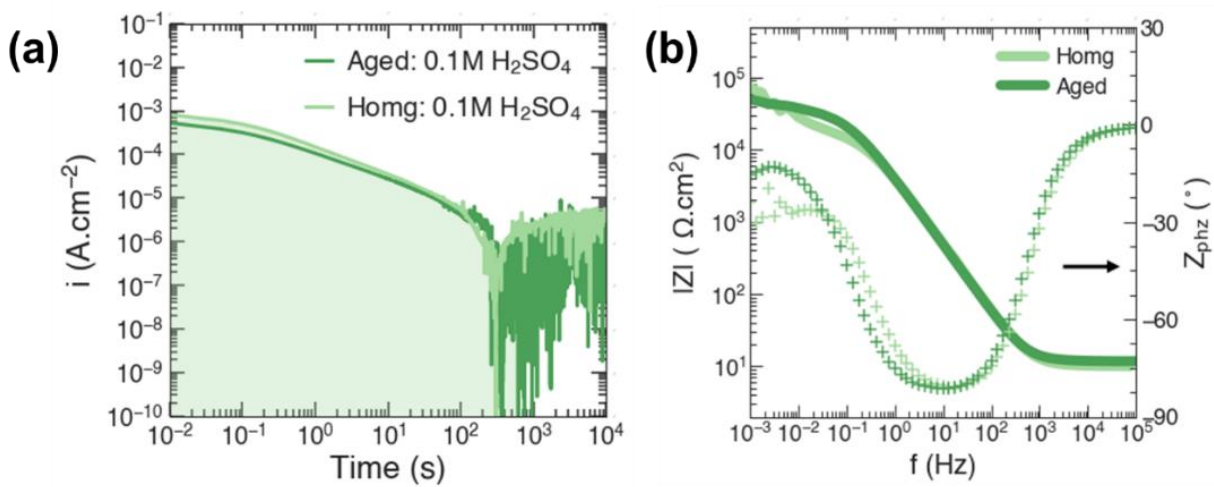

**Supplementary Fig6.** (a) Current decay during the potentiostatic hold at +0.1 V vs. SHE for 10 ks after cathodically reducing the native oxides, and (b) Potentiostatic EIS Bode plots after the 10 ks hold at +0.1 V vs. SHE of as-homogenized and aged CoCrNi alloys, in deaerated chloride free 0.1 mol/L $H_2SO_{4(aq)}$. Fig S2 shows the potentiostatic current decay at 0.1 +V vs. SHE for 10 ks and potentiostatic EIS after the hold at +0.1 V vs. SHE of ordered (as-homogenized CoCrNi) and clustered (aged CoCrNi) in chloride free 0.1 M $H_2SO_4$. The same protocol of three cathodic reduction holds was followed as described in the main text's methods section. The shaded region of first 600s were used to calculate the charge density used to calculate number of dissolved monolayers $h$ for both the alloys. It can be observed that both the alloys show similar behavior of current decays all throughout the period of hold, and impedance performance after 10 ks. This can be attributed to concentration of Cr being much higher than the percolation threshold conc of 13 at% Cr for fcc CoCrNi alloys in chloride free 0.1 M $H_2SO_4$. [11]

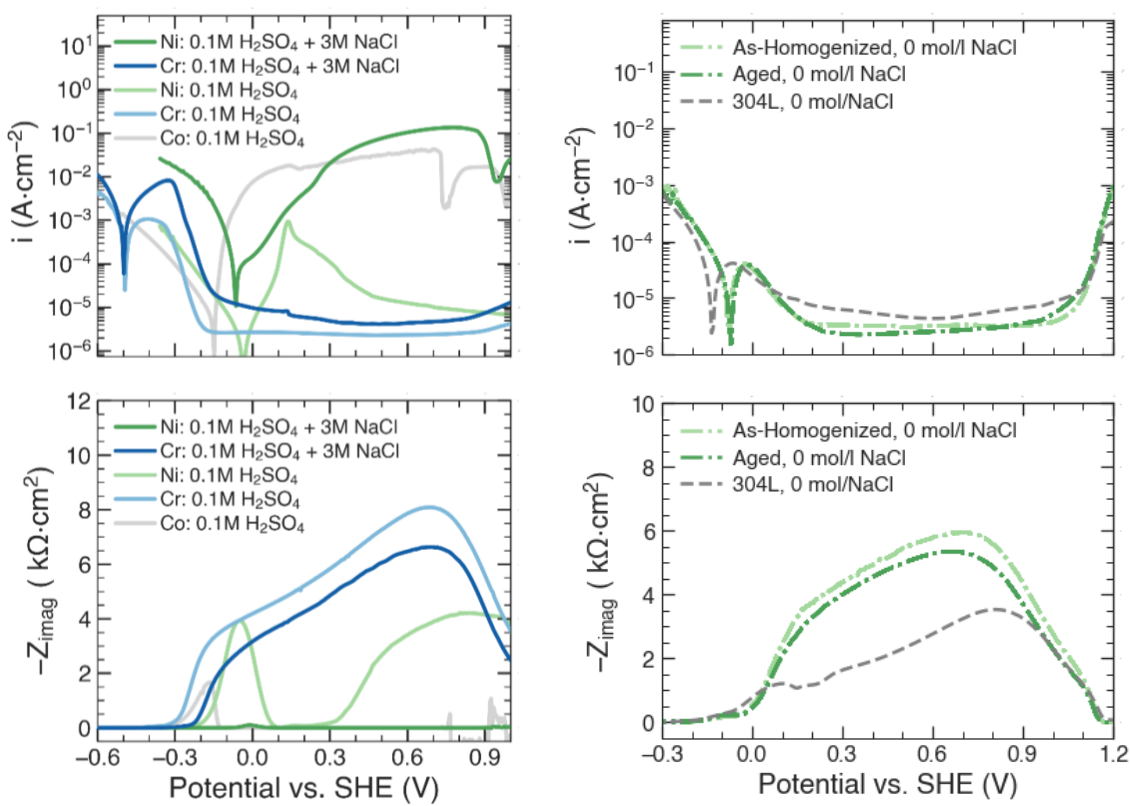

**Supplementary Fig7.** Anodic single frequency linear sweep voltammetry behavior showing current density and $-Z_{imag}$ (monitored in-situ) of : (left) pure Ni, pure Cr, and pure Co in deaerated chloride-free 0.1 mol/L $H_2SO_4$, and that of pure Ni and pure Cr in deaerated 0.1 mol/L $H_2SO_4$ + 3 mol/L NaCl after cathodically reducing their native oxides. (Right): 304L and CoCrNi alloys in chloride free 0.1 mol/l $H_2SO_4$.

Anodic LSV behaviors of Pure Co, Pure Cr, and Pure Ni in 0.1 mol/L $H_2SO_4$ and that of Pure Cr and Pure Ni in 0.1 mol/L $H_2SO_4$ + 3 mol/L NaCl is shown in **Fig S7**. These were obtained after cathodically reducing their native oxides at -1.26 V vs. SHE for 600 s and then polarizing from -0.6 V vs. SHE to +1.0 V vs. SHE with a scan rate of $\approx 0.52\ mV/s$ while *in-situ* monitoring the $-Z_{img}$ with an AC signal of 20 mV(RMS) and 1 Hz. For pure Co and Ni the polarization range was -0.36 V vs. SHE to +1.0 V vs. SHE. Pure Cr passivates around -0.3 V vs. SHE and -0.2 V vs. SHE in with and without 3 mol/L NaCl, respectively as can be seen from current density as well as large $-Z_{img}$ at 1 Hz. Unlike pure Cr, pure Ni passivates in chloride free 0.1 mol/L $H_2SO_4$ around +0.15 V but actively dissolves for anodic potential in 0.1 mol/L $H_2SO_4$ + 3 mol/L NaCl solution. Pure Co which is worse than Ni [32], shows no passivation for any anodic potential and actively dissolves in chloride free 0.1 mol/L $H_2SO_4$ and can be expected to have no passivation rather much higher active dissolution rates in 0.1 mol/L $H_2SO_4$ + 3 mol/L NaCl. Thus, in 3 mol/L NaCl containing acid solution only Cr of the CoCrNi alloy is expected to show passivation to its $Cr_2O_3$ at all anodic potentials.

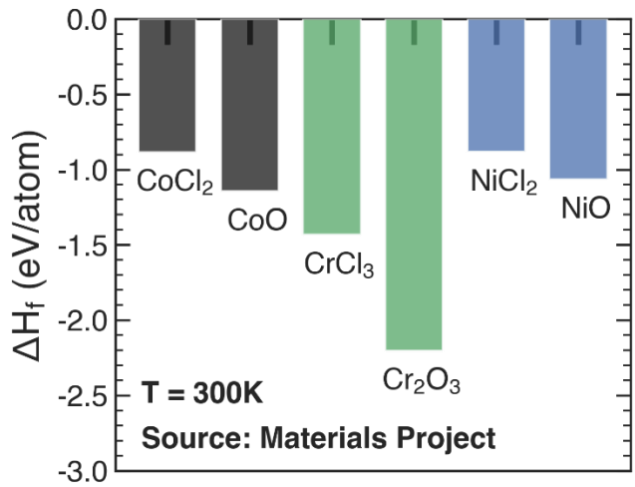

**Supplementary Fig8.** The formation energy of oxides and chloride species of Co, Cr, and Ni at 300 K using Materials Project.

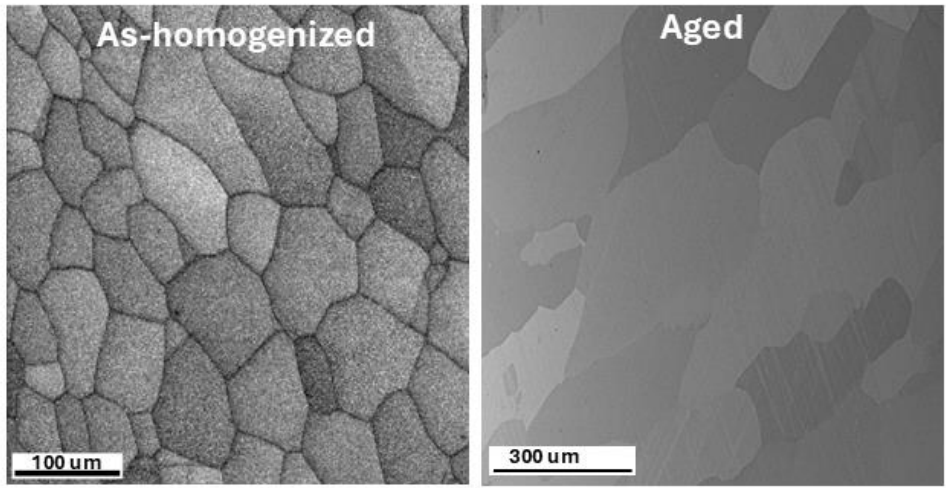

**Supplementary Fig.8.** Microstructure of the as-homogenized (left) and the aged (right) CoCrNi, imaged using scanning electron microscopy. Grain size was measured using ImageJ by thresholding grayscale micrographs and applying the Analyze Particles function to extract individual grain areas. The equivalent circular diameter for each grain was calculated using the formula D=2√A/π with an average taken from 45–50 grains per condition. The measured grain size for the as-homogenized sample was **60.9 ± 2.4 μm**, and for the aged sample was **85.6 ± 2.2 μm**.

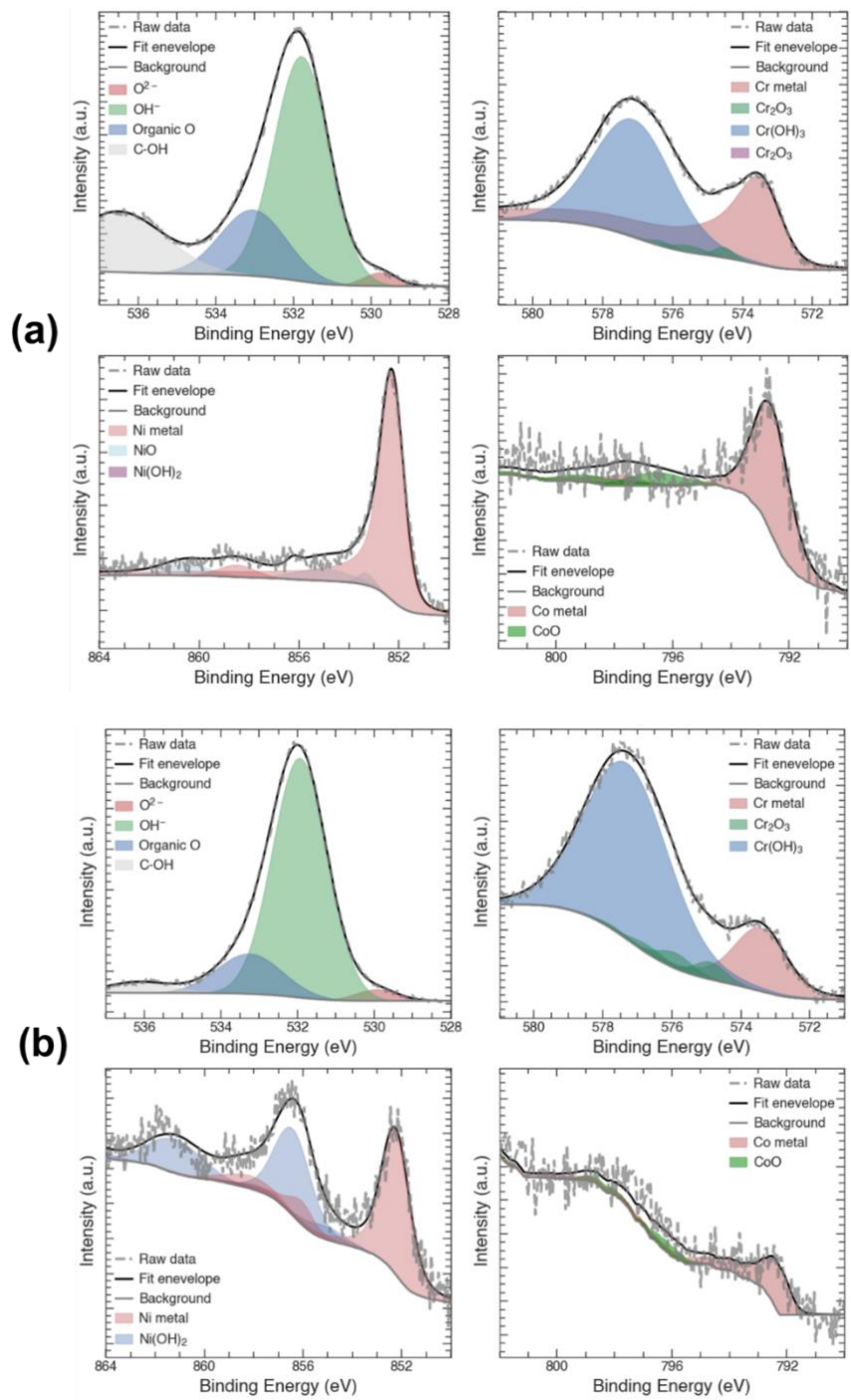

**Supplementary Fig 9.** XPS fit figures for one of the spectra of the samples (a) As-Homogenized, and (b) Aged.

**Correlating oxide thickness and $Z_{imag}$**

Assuming passive films as a parallel plate capacitor, the imaginary component of the impedance $Z_{imag}$ can be directly related to the thickness ($l_{ox}$) of the passive film using the following relation[A].

$$-Z_{imag} = \frac{1}{j\omega C} = \frac{l_{ox}}{j\omega \varepsilon_r \varepsilon_0}$$

Here, $l_{ox}$ is the oxide thickness, $\varepsilon_r$ is the relative permittivity of the oxide, $\varepsilon_0$ is the dielectric constant, $\omega$ is the frequency in radians ($\omega = 2\pi$ as $f = 1$ Hz for the single-frequency LSV experiments).

## Atomic Simulation

The Large-Scale Atomic / Molecular Massively Parallel Simulator (LAMMPS)[B] is used to preform hybrid Monte Carlo - Molecular Dynamics (MC-MD) simulations under the canonical ensemble to [C-F] equilibrate microstructures to a target composition of 33.33-at% Ni, 33.33-at% Co, and 33.33-at% Cr. Two independent MC-MD simulations are performed at finite temperature to produce varying degrees of chemical order. The MC-MD routine is initially applied to an fcc single crystal with $a = 3.56$ A [H] and a chemical ordering equivalent to that of an fcc random solid solution. Each simulation was determined converged after stabilization of both the potential energy and Warren-Cowley parameters. After performing each simulation to convergence at 300°C, the equiatomic alloy develops regions of segregation and clustering. An additional model was equilibrated at an MCMD temperature of 600°C. Favorable ordered pairs in this structure matched that of the 300°C simulations, however the ordering length-scale was smaller. The authors chose to display the 300°C data to illustrate the strength of the ordered pairs that develops at low temperatures, relative to that in an as-homogenized sample. The microstructure of a single microstate of this alloy drawn from the computationally generated ensemble after equilibrating for 1 million MD timesteps at each temperature. For every 100 MC swap attempts, 10,000 MD steps are performed under NPT conditions. The MD timestep was set to 2.5 fs and the dimensions of the fcc lattice were approximately 6.1 nm x 5 nm x 5.8 nm, resulting in a simulation cell of 16,000 atoms.

## Short-Range Order Calculations

Warren-Cowley Short-Range Order parameters (WCP) [I] are used to describe the nature of chemical ordering in each microstructure. These are correlation parameters which assign a value to the pairwise bonding tendencies within a system. The formal expression for this parameter can be seen in the equation below. In this study, a LAMMPS compute [J] was employ to analyze each atom's local chemical environment. A purely random solid solution is denoted by a value of zero, whereas positive values correspond to a tendency for segregation and negative values indicate ordering.

$$\alpha_{ij}^n = 1 - \frac{P_{ij}^n}{\delta_{ij} - c_j}$$

The value of n corresponds to the nth nearest neighbor shell for the reference atom (*i*), with the probability ($P_{ij}^n$) of having a neighbor of type (*j*) in that shell. The concentration ($c_j$) of (*j*) in the system is used, as well as the Kronecker delta function ($\delta_{ij}$). This function is equal to 1 if *i=j*, and zero otherwise. These correlation parameters are used to uncover the composition and magnitude of the ordering events in a system by assigning a value to each type of pairwise bond. Larger values correspond to a larger interaction between the species, the sign of which is dependent on the tendency for repulsion or attraction.

***Supplementary table 1.*** *Bond length and Debye-Waller factors of Ni, Co, and Cr with their nearest neighbor atoms in as-homogenized and aged CoCrNi alloys from the EXAFS fitting measurements.*

| Absorption edge | | $\sigma^2$ (Å) | $\Delta R$ (Å) | $R_0$ (Å) | $R_i$ (Å) |
|---|---|---|---|---|---|
| **As-homogenized alloy** | Co | 0.0061(8) | 0.005 (6) | 2.491 | **2.496** |
| | Cr | 0.0064(7) | -0.055(5) | 2.552 | **2.497** |
| | Ni | 0.0066(5) | -0.008(4) | 2.495 | **2.494** |
| **Aged alloy** | Co | 0.0061(9) | -0.001 (6) | 2.491 | **2.490** |
| | Cr | 0.0049(6) | -0.059(4) | 2.552 | **2.493** |
| | Ni | 0.0071(7) | -0.005(5) | 2.495 | **2.490** |

***Supplementary table 2.*** *Calculated interatomic distances in pure Co, Cr, and Ni [K].*

| *Element* | Atomic radius(nm) | Interatomic distance (Å) |
|---|---|---|
| *Co* | 0.1251 | **2.502** |
| *Cr* | 0.1249 | **2.498** |
| *Ni* | 0.1245 | **2.490** |

## Supplementary Materials References

[A] B.Hirschorn, M.E.Orazem, B.Tribollet, V. Vivier, I. Frateur, and M. Musiani, "Determina tion of effective capacitance and film thickness from constant-phase-element parameters," Electrochimica Acta, vol. 55, no. 21, pp. 6218–6227, Aug. 2010. DOI: 10.1016/j.electacta.2009.10.065.

[B] A.P. Thompson, H.M. Aktulga, R. Berger, D.S. Bolintineanu, W.M. Brown, P.S. Crozier, P.J. in't Veld, A. Kohlmeyer, S.G. Moore, T.D. Nguyen, LAMMPS-a flexible simulation tool for particle-based materials modeling at the atomic, meso, and continuum scales, Comput. Phys. Commun. 271 (2022) 108171.

[C] B. Sadigh, P. Erhart, A. Stukowski, A. Caro, E. Martinez, L. Zepeda-Ruiz, Scalable parallel Monte Carlo algorithm for atomistic simulations of precipitation in alloys, Phys. Rev. B. 85 (2012) 184203.

[D] S. Plimpton, Fast parallel algorithms for short-range molecular dynamics, J. Comput. Phys. 117 (1995) 1–19.

[E] B. Sadigh, P. Erhart, Calculation of excess free energies of precipitates via direct thermodynamic integration across phase boundaries, Phys. Rev. B. 86 (2012) 134204.

[F] A. Stukowski, B. Sadigh, P. Erhart, A. Caro, Efficient implementation of the concentration-dependent embedded atom method for molecular-dynamics and Monte-Carlo simulations, Model. Simul. Mater. Sci. Eng. 17 (2009) 75005.

[G] J.M. Cowley, An approximate theory of order in alloys, Phys. Rev. 77 (1950) 669–675.

[H] B.C. Sales, K. Jin, H. Bei, G.M. Stocks, G.D. Samolyuk, A.F. May, M.A. McGuire, Quantum critical behavior in a concentrated ternary solid solution, Sci. Rep. 6 (2016) 26179.

[I] Cowley, J. M. An Approximate Theory of Order in Alloys. *Phys. Rev. B.* 77 (5) (1950).

[J] Antillon, E., Woodward, C., Rao, S. I., Akdim, B., & Parthasarathy, T. A. (2020). Chemical short range order strengthening in a model FCC high entropy alloy. *Acta Materialia*, *190*, 29–42.

[K] Zhang, Y., Zuo, T. T., Tang, Z., Gao, M. C., Dahmen, K. A., Liaw, P. K., & Lu, Z. P. (2014). Microstructures and properties of high-entropy alloys. *Progress in Materials Science, 61*, 1–93. https://doi.org/10.1016/j.pmatsci.2013.10.001